\documentclass[aps,twocolumn,prl]{revtex4}

\usepackage{amsmath}
\usepackage{amssymb}
\usepackage{graphicx}
\usepackage{bm}

\begin{document}

\title{Geometric phase of a moving dipole under a magnetic field at a distance}%
\author{Kang-Ho Lee}
\author{Young-Wan Kim}
\author{Kicheon Kang}
\email{kicheon.kang@gmail.com}
\affiliation{Department of Physics, Chonnam National University,
 Gwangju 500-757, Korea}
\date{\today}

\begin{abstract}
We predict a geometric quantum phase shift of a moving electric dipole 
in the presence of an external
magnetic field at a distance. 
On the basis of the Lorentz-covariant field interaction approach, 
we show that a geometric phase appears under the condition 
that the dipole is moving in the field-free region,
which is distinct from the topological He-McKellar-Wilkens phase 
generated by a 
direct overlap of the dipole and the field. 
We discuss the experimental 
feasibility of detecting this phase with atomic interferometry 
and argue that detection of this 
phase would result in a deeper understanding of the locality in quantum
electromagnetic interaction. 
\end{abstract}


\maketitle
\newcommand \tr {{\rm Tr}}

{\em Introduction.}-
A moving charge under an external electric/magnetic field exhibits a 
topological quantum phase shift known as the Aharonov-Bohm~(AB) 
effect~\cite{aharonov59}. An intriguing aspect of the AB effect is
that it takes place even without a local overlap between the charge and the
external field~(see Ref.~\onlinecite{peshkin89} for a review).
A similar topological phase shift, namely the He-McKellar-Wilkens (HMW) 
phase~\cite{he93,wilkens94},
was predicted for an electric dipole under an external magnetic field.
The HMW phase shift is given by
\begin{equation}
 \Delta\phi_{HMW} = \frac{1}{\hbar c} \oint \mathbf{B}\times\mathbf{d}
  \cdot d\mathbf{R}  
\label{eq:hmw}
\end{equation}
for a closed path $\mathbf{R}$ of the dipole, 
where $\mathbf{B}$ and $\mathbf{d}$ are the external magnetic field and
the dipole moment, respectively. This phase can be derived
from the interaction Lagrangian of a moving dipole under 
a magnetic field, introduced by
Wilkens~\cite{wilkens94}:
\begin{equation}
 L_W = 
    \frac{1}{c} \mathbf{d}\cdot(\dot{\mathbf{R}}\times\mathbf{B}) \,.
\label{eq:LW}
\end{equation}
%
Despite their similarity,  there is a notable difference between
the HMW and the AB phases.
For an observation of the HMW phase, the moving 
particle should experience a direct overlap with the external magnetic field.
In contrast, the AB phase shift
can be generated without a direct overlap of the interfering particle and the  
external field. 
An interesting question here is
whether any measurable result arises from the interaction
between a dipole and the external field without a direct 
overlap of the two entities. 
As far as the topological phase is concerned, no measurable
effect is produced between the 
nonoverlapping $\mathbf{d}$ and $\mathbf{B}$, 
which is apparent from Eqs.~\eqref{eq:hmw} and \eqref{eq:LW}. 
This point was addressed 
in Refs.~\onlinecite{spavieri98} and \onlinecite{wilkens98}.

In this Letter, we predict a novel quantum phase shift of
a moving dipole under a magnetic field in the absence of a direct overlap of
the two entities. 
Our prediction is based on the 
Lorentz-covariant field-interaction (LCFI) approach~\cite{kang13,kang15}. 
The essence of the LCFI theory is that, the effect of an external
electromagnetic field on a moving charge can be universally represented 
by the interaction between the two electromagnetic fields,   
the external field and that generated by the charge,
along with the incorporation of the relativity principle. 
The electromagnetic potential is eliminated accordingly.
It was demonstrated that the LCFI approach reproduces the 
established result for the topological AB effect. In addition, 
this theory leads to the prediction of a new type of quantum interference 
that cannot be described
in the standard potential-based theory~\cite{kang15a}. 
Here, on the basis of the LCFI framework, we
find a non-topological geometric
quantum phase of a moving dipole in the absence of
the local overlap of $\mathbf{d}$ and $\mathbf{B}$. 
We also propose a realistic setup for observing this phase with 
a hydrogenlike atom 
under a distant semi-infinite magnetic sheet. 
Notably, this geometric phase cannot be predicted with either the Wilkens'
(Eq.~\eqref{eq:LW}) or 
the standard potential-based approaches. 
This implies that 
a successful experiment would shed light on our
understanding of the quantum locality and the role of potential 
in the electromagnetic interaction.

{\em Lagrangian and Hamiltonian for an electric dipole.}-
In the LCFI approach,
the influence of the external
electric~($\mathbf{E}$) and magnetic~($\mathbf{B}$) fields on 
a moving charge $q$ is described by the Lorentz-covariant
interaction Lagrangian~\cite{kang13,kang15}
\begin{equation}
 L_{int}^q = \frac{1}{4\pi} \int ( 
       \mathbf{B}_q\cdot\mathbf{B} - \mathbf{E}_q\cdot\mathbf{E})d^3\mathbf{r}' 
       \,,
\label{eq:Lint_EB}
\end{equation}
where $\mathbf{E}_q$($\mathbf{B}_q$) represents the electric(magnetic)
field generated by the charge $q$.
It was previously shown that this Lagrangian reproduces the
force-free topological AB effect, which is equivalent to the well established
result obtained from the potential-based theory~\cite{kang13,kang15}.
This demonstrates the validity of the field-interaction framework. 
Here, we deal with the quantum effect 
generated by the external $\mathbf{B}$
and drop the second term in the right-hand side of 
Eq.~\eqref{eq:Lint_EB}~\cite{note-eom}.
Eq.~\eqref{eq:Lint_EB} can then be rewritten as
\begin{subequations}
\label{eq:Lint2}
\begin{equation}
 L_{int}^q = \dot{\mathbf{r}} \cdot \mathbf{\Pi}_q \,,
\end{equation}
where $\dot{\mathbf{r}}$ is the velocity of charge $q$ and
\begin{equation}
 \mathbf{\Pi}_q = \frac{1}{4\pi c} \int \mathbf{E}_q \times 
 \mathbf{B} \  d^3 \mathbf{r}'
\label{eq:Pi_q}
\end{equation}
\end{subequations}
is the field momentum produced by the overlap between
$\mathbf{E}_q$ and $\mathbf{B}$. 

The LCFI approach is not limited to a single point charge and can be
generalized to an arbitrary distribution of charges. 
We adopt it for describing the interaction between the
dipole and the external field.
Applying Eq.~\eqref{eq:Lint2}, 
the interaction between an electric dipole (composed of two point charges 
$q$ and $-q$) 
and $\mathbf{B}$ is described by the interaction Lagrangian 
\begin{equation}
 L_{int} = \dot{\mathbf{r}}_q \cdot  \mathbf{\Pi}_q(\mathbf{r}_q)
         + \dot{\mathbf{r}}_{\bar{q}} \cdot  \mathbf{\Pi}_{\bar{q}}
                   (\mathbf{r}_{\bar{q}}) \,,
\end{equation}
where $\mathbf{r}_q$($\mathbf{r}_{\bar{q}}$) and 
$\mathbf{\Pi}_q$($\mathbf{\Pi}_{\bar{q}}$) are the position of 
charge $q$ ($-q$) and the corresponding field momentum defined
in Eq.~\eqref{eq:Pi_q}, respectively. Applying the dipole approximation 
\begin{displaymath}
 \mathbf{\Pi}_q(\mathbf{r}_q) \simeq \mathbf{\Pi}_q(\mathbf{R}) 
   + (\mathbf{r}_q-\mathbf{R})\cdot\nabla\mathbf{\Pi}_q(\mathbf{R}), 
\end{displaymath}
we find 
\begin{equation}
  L_{int} = \dot{\mathbf{R}}\cdot(\mathbf{r}\cdot\nabla)
            \mathbf{\Pi_{q}}(\mathbf{R})   
          + \dot{\mathbf r}\cdot\mathbf{\Pi_{q}}(\mathbf {R}), 
\label{eq:Lint_a}
\end{equation}
where $\mathbf{R}$ and $\mathbf{r} \equiv \mathbf{r}_q - \mathbf{r}_{\bar{q}}$
are the center of mass and the relative coordinates of the two charges, 
respectively.
Here the gradient $\nabla$ is associated with the $\mathbf{R}$ coordinates;
that is, $\nabla\equiv\nabla_{\mathbf{R}}$.

The field momentum $\mathbf{\Pi_{q}}$ satisfies the relation
\begin{equation}
 \nabla\times\mathbf{\Pi}_q = \frac{q}{c} \mathbf{B} \,.
\end{equation}
Together with the vector identity, this relation leads to an instructive form of
$L_{int}$:
\begin{equation}
L_{int}  = \frac{q}{c}\dot{\mathbf R}\cdot\mathbf{B}\times\mathbf{r}
         + \dot{\mathbf R}\cdot\nabla(\mathbf{r}\cdot\mathbf{\Pi}_q)
         + \dot {\mathbf {r}} \cdot \mathbf{\Pi}_q  \,.
\label{eq:Lint_b}
\end{equation}
The first term of the right-hand side in Eq.~\eqref{eq:Lint_b} is equivalent to
the Wilkens' Lagrangian of Eq.~\eqref{eq:LW}.
For a time-independent $\mathbf{B}$, one can find the relation
\begin{equation}
 L_{int} = L_W + \frac{d}{dt} (\mathbf{r}\cdot\mathbf{\Pi}_q) \,,
\end{equation}
which  shows that
the two Lagrangians, $L_W$ and $L_{int}$, predict the same
topological HMW phase shift of Eq.~\eqref{eq:hmw} 
together with the same classical
equation of motion. Therefore, $L_{int}$ also provides a vanishing HMW phase 
in the absence of a direct overlap between $\mathbf{d}$ and
$\mathbf{B}$.   
It is often believed that a total time derivative
in a Lagrangian does not produce any physical effect, but this is not true in 
general 
(see Ref.~\onlinecite{kang15a} for details).
The equivalence of $L_W$ and $L_{int}$ is limited to classical phenomena
and topological quantum phases.

The net Lagrangian of an electric dipole under an external magnetic field
is written as
\begin{equation}
 L = \frac{1}{2}M\dot{\mathbf R}^2 + \frac{1}{2}m\dot{\mathbf r}^2
   - V(\mathbf {r}) + L_{int} \,,
\end{equation}
with $L_{int}$ given by
Eq.~\eqref{eq:Lint_b}. 
$M$, $m$ and $V(\mathbf {r})$ are the total mass, 
reduced mass, and the scalar potential describing the
interaction between $q$ and $-q$, respectively.
A standard Legendre transformation provides the corresponding Hamiltonian, 
\begin{subequations}
\begin{equation}
 H = H_R + H_r \,,
\end{equation}
where $H_R$($H_r$) represents the term associated with 
the dipole's center of motion (internal degree of freedom) 
given by
\begin{eqnarray}
 H_R &=&  \frac{1}{2M} \left[ \mathbf P 
         -\frac{q}{c}\dot{\mathbf R}\cdot\mathbf{B}\times\mathbf{r}
         -\nabla (\mathbf{r}\cdot\mathbf{\Pi}_q)
                      \right]^2 , \label{eq:H_R} \\
 H_r &=&  \frac{1}{2m}(\mathbf p-\mathbf {\Pi}_q)^2+V(\mathbf{r}) \,.
         \label{eq:H_r}
\end{eqnarray}
\end{subequations}
Here $\mathbf{P}=\partial{L}/\partial{\dot{\mathbf{R}}}$ and 
$\mathbf{p}=\partial{L}/\partial{\dot{\mathbf{r}}}$ are the canonical momenta 
of the coordinates $\mathbf{R}$ and $\mathbf{r}$, respectively.

{\em Appearance of a geometric phase.}-
We focus on the case without a direct overlap between 
the dipole and $\mathbf{B}$, 
and drop the corresponding term (represented by
$\frac{q}{c}\dot{\mathbf R}\cdot\mathbf{B}\times\mathbf{r}$ in 
$H_R$ of Eq.~\eqref{eq:H_R}). In this case, the topological HMW phase is
absent,  and the effect of the external
$\mathbf{B}$ is manifested by the field momentum $\mathbf{\Pi}_q$.

$\mathbf{\Pi}_q$ appears in both $H_r$ and $H_R$ of the Hamiltonian $H$. 
First, the $\mathbf{\Pi}_q$ term in $H_r$~(Eq.~\eqref{eq:H_r}) 
has no physical effect, because 
$\mathbf{\Pi}_q$ is a function of $\mathbf{R}$ alone.
To show this explicitly, let us set $\mathbf{\Pi}_q = \Pi_q\hat{z}$ 
($\hat{z}$ being the unit vector along the $z$-axis)
without loss of generality.
Then one can find that
\begin{subequations}
\begin{equation}
 H_r = 
  \frac{1}{2m}(\mathbf p-\nabla_\mathbf{r}\chi(\mathbf{r}))^2+V(\mathbf{r}) \,,
\end{equation}
where 
\begin{equation}
 \chi(\mathbf{r}) = \Pi_q z  \,.
\end{equation}
\end{subequations}
This introduces only
an overall phase factor of $\exp{(i\Pi_qz)}$ for any internal state. 
This factor is a single-valued function and thus does not provide
any measurable consequence.
Therefore, the internal dynamics of the dipole is independent
of $\mathbf{B}$, and the effect of the interaction with $\mathbf{B}$
is manifested in
\begin{equation}
 H_R =  \frac{1}{2M} \left[ \mathbf P 
         -\frac{1}{q}\nabla (\mathbf{d}\cdot\mathbf{\Pi}_q)
                      \right]^2  \,.
\label{eq:H_R2}
\end{equation}
Here the $\mathbf{r}$-coordinate is replaced by the dipole variable 
$\mathbf{d}$($=q\mathbf{r}$).
As will be shown below, our main finding is that
a measurable
phase shift appears owing to the interaction term 
of $\nabla(\mathbf{d}\cdot\mathbf{\Pi}_q)/q$,
despite the absence of the direct overlap of $\mathbf{d}$ and $\mathbf{B}$.

According to the Hamiltonian of 
Eq.~\eqref{eq:H_R2}, a freely moving dipole from $\mathbf{R}=\mathbf{R}_i$ to 
$\mathbf{R}=\mathbf{R}_f$ acquires a phase shift of
\begin{equation}
 \phi = \frac{1}{\hbar q}
        \int_{\mathbf{R}_i}^{\mathbf{R}_f} 
             \nabla(\mathbf{d}\cdot\mathbf{\Pi}_q)\cdot d\mathbf{R}
      = \frac{1}{\hbar q} [
                     \mathbf{d}\cdot\mathbf{\Pi}_q(\mathbf{R}_f)
                   - \mathbf{d}\cdot\mathbf {\Pi}_q(\mathbf{R}_i)
                          ].
\label{eq:phase-dipole}
\end{equation}
To be specific, let us consider a hydrogenlike atom under a distant
semi-infinite magnetic sheet (Fig.~1) represented by 
\begin{equation}
 \mathbf B(\mathbf{r}') = 
   \left\{ \begin{array}{ll} 
       B_0\hat{x} & \mbox{\rm for } z'\geq 0 \;\; \mbox{\rm and }\; 
                                     -y_0 \leq y' < 0 ,
  \\
       0 & \mbox{\rm otherwise} . 
            \end{array}
    \right.
\label{eq:B}
\end{equation}
The atom is moving along the $z$-axis 
from $\mathbf{R}_i = a\hat{y} + z_i\hat{z}$
to $\mathbf{R}_f = a\hat{y} + z_f\hat{z}$,
a path with $\mathbf{B}=0$. 
%
The field momentum $\mathbf{\Pi}_q$ can be evaluated from the definition
in Eq.~\eqref{eq:Pi_q} together with $\mathbf{B}$ in Eq.~\eqref{eq:B} 
and
\begin{equation}
 \mathbf{E}_q(\mathbf{r}')
    = q(\mathbf{r}'-\mathbf{R})/|\mathbf{r}'-\mathbf{R}|^3 ,
\end{equation}
at the position $\mathbf{r}'$. (Note that the relativistic
correction is negligible in the limit of $|\dot{\mathbf{R}}|\ll c$.)
Evaluation of $\phi$ requires the values of $\mathbf\Pi_q$ 
at the two end points, 
$\mathbf{R} = \mathbf{R}_i$
and $\mathbf{R} = \mathbf{R}_f$, as $\phi$ depends only on
the initial and the final values of $\mathbf\Pi_q$ 
(see Eq.~\eqref{eq:phase-dipole}). 
Under the condition $z_i\ll -a$ and 
$z_f\gg a$, we obtain
\begin{equation}
 \mathbf\Pi_q(\mathbf{R}_i) = 0, \;\; \mbox{\rm and} \;\;
 \mathbf\Pi_q(\mathbf{R}_f) = \frac{n_Bq}{2c}\hat{z} ,
\end{equation}
where $n_B\equiv B_0y_0$ is the linear density of the magnetic flux along
the $z$-axis.
Therefore, the phase shift is found to be
\begin{equation}
 \phi = \frac{n_Bd_z}{2\hbar c} \,,
\end{equation}
where $d_z$ is the $z$-component of the dipole moment $\mathbf{d}$.

Detection of this phase requires a superposition of different dipole
states, as the phase depends on $d_z$. 
This can be achieved, for example, 
in the excited states of the hydrogenlike atom~(see e.g., 
Ref.~\onlinecite{gasiorowicz07}).
For $2s$ and $2p$ states of the hydrogen atom, $|2lm\rangle$, where
$l\in \{0,1\}$ and $m\in \{-1,0,1\}$, 
the dipole eigenstates are 
\begin{equation}
 |\pm\rangle = \frac{1}{\sqrt{2}} \left( |200\rangle \pm |210\rangle 
                                    \right), 
\end{equation} 
with their eigenvalues $d_z = \pm 3ea_0$
($a_0$ being the Bohr radius).
A possible procedure for the proposed experiment is as follows. 
Initially, the atom is prepared in the state
\begin{equation}
 |\psi_0\rangle = |200\rangle 
  = \frac{1}{\sqrt{2}} \left( |+\rangle + |-\rangle \right)
 \,
\end{equation}
at $\mathbf{R}=\mathbf{R}_i$.
The atom moves parallel to the magnetic sheet, and the evolution of the
state is influenced by the interaction
with the external field (as shown in Eq.~\eqref{eq:H_R2}). Upon arrival at
$\mathbf{R} = \mathbf{R}_f$, the state evolves into
(neglecting the overall dynamical phase factor)
\begin{subequations}
\begin{eqnarray}
 |\psi_f\rangle &=& \frac{1}{\sqrt{2}} 
     \left( |+\rangle e^{i\phi_g} + |-\rangle e^{-i\phi_g} 
     \right)  \nonumber \\
  &=& \cos{(\phi_g)}|200\rangle 
                 + i\sin{(\phi_g)}|210\rangle \,,
\end{eqnarray}
where 
\begin{equation}
 \phi_g = 3ea_0n_B/2\hbar c \,.
\label{eq:phi_g}
\end{equation}
\label{eq:psi_f}
\end{subequations}
That is, a measurable geometric phase shift $\phi_g$ appears 
as a result of the the interaction with
the magnetic field at a distance. 
The magnitude of $\phi_g$ is given by
\begin{equation}
 \phi_g = 1.205\times 10^{-1} \mbox{\rm (Gauss}\cdot \mbox{\rm cm)}^{-1} n_B ,
\end{equation}
a realistic value that can be achieved and controlled experimentally.

{\em Discussion.}- 
Let us discuss the properties of the geometric phase 
$\phi_g$ in more detail.
First, as described above, $\phi_g$ appears without 
a direct overlap of the
dipole ($\mathbf{d}$) and the external magnetic field ($\mathbf{B}$). 
This is in contrast with the
HMW phase where a topological phase builds up as a result of the overlap
of the two quantities.
In addition, the absence of the overlap implies 
that the force does not play any role in the appearance of $\phi_g$.

Second, the Maxwell dual phase of $\phi_g$ in Eq.~\eqref{eq:phi_g} 
is automatically predicted due to the 
invariance of the Maxwell equations 
under the electric-magnetic duality transformations~\cite{jackson62}. 
Maxwell equations are invariant under the 
transformations given by $\mathbf{E}\rightarrow\mathbf{B}$,
$\mathbf{d}\rightarrow\vec{\mu}$, $\mathbf{B}\rightarrow -\mathbf{E}$,
and $\vec{\mu}\rightarrow -\mathbf{d}$, where $\vec{\mu}$ stands for the
magnetic dipole moment vector. This invariance leads to the prediction that
a magnetic moment with an electric field at a distance would undergo a
similar phase shift.
Note that this Maxwell duality was adopted for elucidating the relations
between the topological AB, Aharonov-Casher~\cite{aharonov84}, and the HMW 
phases~\cite{dowling99}.
Let us consider the setup illustrated in Fig.~2, which is the
Maxwell dual of the one in Fig.~1. 
A spin $1/2$ particle (magnetic dipole) is moving along the $z$-axis under
a distant semi-infinite sheet of the electric field. 
The initial state is prepared as 
\begin{equation}
 |\psi_0\rangle = \frac{1}{\sqrt{2}} 
      \left( |\uparrow\rangle + |\downarrow\rangle \right) \,,
\end{equation}
where $|\uparrow\rangle$ ($|\uparrow\rangle$) 
represents the spin up (down) state in the $z$-axis.
The particle interacts with the distant electric field via the Maxwell dual
interaction of $L_{int}$ in Eq.~\eqref{eq:Lint_b}, 
and the state evolves into 
\begin{subequations}
\begin{equation}
 |\psi_f\rangle = \frac{1}{\sqrt{2}} 
      \left( e^{i\phi_g^m} |\uparrow\rangle + 
             e^{-i\phi_g^m}  |\downarrow\rangle \right) \,,
\end{equation}
where $\phi_g^m$, the Maxwell dual of $\phi_g$ in 
Eq.\eqref{eq:phi_g}, is found to be
\begin{equation}
 \phi_g^m = -\frac{n_E\mu}{2\hbar c} \,.
\end{equation}
\end{subequations}
Here, $n_E$ and $\mu$ stand for the linear density of the electric flux
and the magnetic moment of the moving particle, respectively.
The physical values of $\phi_g^m$ are estimated to be 
$\phi_g^m = 5.1\times 10^{-10}n_E\mu/
(\mbox{\rm Volt})$ and $\phi_g^m\sim {\cal O} (10^{-6})n_E/(\mbox{\rm Volt})$ for a neutron and for
an atom, respectively, and can be experimentally achieved.

Finally, our study strongly supports the locality of the 
electromagnetic interaction, which in the present 
case is manifested in the local 
interaction between 
the external magnetic field and that produced by
the dipole charges.
To show this, let us check the prediction of $\phi_g$ 
in the standard potential-based approach. In this approach,
$H_R$ in Eq.~\eqref{eq:H_R} 
is replaced by~\cite{spavieri98}
\begin{equation}
 H_R =  \frac{1}{2M} \left[ \mathbf P 
         -\frac{1}{c}\dot{\mathbf R}\cdot\mathbf{B}\times\mathbf{d}
         -\frac{1}{c}\nabla (\mathbf{d}\cdot\mathbf{A})
                      \right]^2  \,.
\label{eq:H_RA}
\end{equation} 
The phase shift of the moving dipole (for the state in Eq.~\eqref{eq:psi_f})
evaluated from this Hamiltonian is
\begin{equation}
 \phi_g = \frac{3ea_0}{\hbar c}
          \left[ A_z(\mathbf{R}_f) - A_z(\mathbf{R}_i)
          \right] \,.
\end{equation}
That is, $\phi_g$ is proportional to the difference between
the $z$-components of the
vector potential at the two end points. 
However, this is problematic, as the derived $\phi_g$  
is gauge dependent. It does not imply a prediction of a null 
phase shift, 
but demonstrates the invalidity of the usual potential-based 
picture of the electromagnetic interactions. 
The origin of this problem is the absence of a well-defined local interaction
of gauge-invariant quantities.
Note that Wilkens' Lagrangian
leads to $\phi_g=0$, which is not a real prediction but a 
result of a specific choice of gauge satisfying 
$\nabla (\mathbf{d}\cdot\mathbf{A}) = 0$. Any choice of gauge is fine for
the topological phase (as also noticed in Ref.~\onlinecite{wilkens98}),
but cannot provide a proper description of the non-topological phase
predicted here.
Therefore, successful observation of $\phi_g=3ea_0 n_B/\hbar c$ 
can confirm the validity of the LCFI approach and eliminate the 
gauge-dependent potential picture.
This is remarkable in that
it removes the dynamical nonlocality~\cite{popescu10} in the quantum state 
evolution.
In other words, our study shows that any quantum dynamics
involving electromagnetic interaction can be described in terms of 
locally realistic variables alone.

{\em Conclusion.}- 
We have predicted a geometric phase shift of a moving electric
dipole under an external magnetic field at a distance. 
Prediction of this 
phase is possible with the Lorentz-covariant field-interaction
theory~\cite{kang13,kang15}, whereas the standard approaches fail. 
We have also proposed a realistic experiment with hydrogenlike atoms
under a semi-infinite magnetic sheet. Our result demonstrates that
the LCFI approach to quantum electromagnetic interactions 
opens a new window of discovery that cannot be conceived
in the conventional framework. 

{\em Acknowledgment.} 
This work was supported by the National Research Foundation of Korea (NRF)
(Grant No.~NRF-2015R1D1A1A01057325).

\bibliography{references}

\begin{thebibliography}{15}
\bibitem[{not()}]{note-eom}
\bibinfo{note}{To be precise, it is necessary to maintain both terms of the
  right-hand side of Eq.~\eqref{eq:Lint_EB} to get a correct classical equation
  of motion including the source of the external field, even when
  $\mathbf{E}=0$ in the lab frame. However, it does not affect evaluation of a
  quantum phase. (See discussion in Ref.~\onlinecite{kang13}).}
\end{thebibliography}
%
%

\begin{figure}
 \includegraphics[width=3.4in]{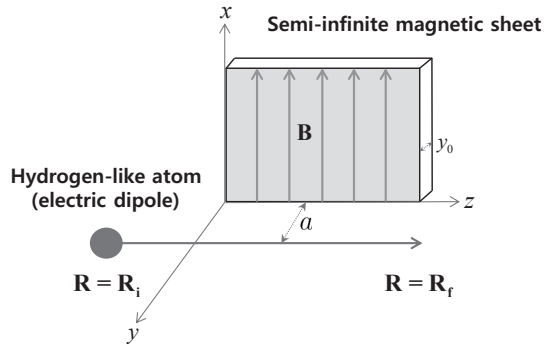}
\caption{Schematic setup for detection of the geometric phase: 
 A hydrogenlike atom is moving parallel to a distant semi-infinite magnetic
 sheet. The accumulated phase shift $\phi_g$ due to the field interaction 
 can be detected by monitoring the final state of the dipole 
 (see Eq.~\eqref{eq:psi_f}). 
 }
\end{figure}
\begin{figure}
 \includegraphics[width=3.4in]{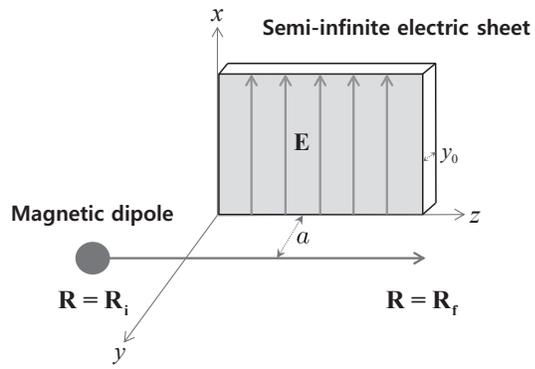}
\caption{Schematic setup for detection of $\phi_g^m$, the
 Maxwell dual of $\phi_g$: 
 A particle with spin $1/2$ is moving parallel to a distant semi-infinite 
 sheet of electric field.
 }
\end{figure}

\end{document}